\documentclass{eas}
\usepackage{graphicx}
\usepackage{mathrsfs}
\usepackage{amssymb}
\usepackage{amsmath}
\usepackage{graphicx}
\usepackage{bm}
\begin{document}

\title{Exclusion, Discovery and Identification of Dark Matter with Directional Detection} 
\runningtitle{Phenomenology of Directional Detection of Dark Matter}
\author{J. Billard}\address{Laboratoire de Physique Subatomique et de Cosmologie, Universit\'e Joseph Fourier Grenoble 1,
  CNRS/IN2P3, Institut Polytechnique de Grenoble, Grenoble, France}
\author{F. Mayet}\sameaddress{1}
\author{D. Santos}\sameaddress{1}
%
%
\begin{abstract}
 Directional detection is a promising search strategy to discover galactic Dark Matter.  
We present a Bayesian analysis framework dedicated to data from upcoming directional detectors. 
The interest of directional detection as a powerful tool to set exclusion limits, to authentify a Dark Matter detection or to constrain the Dark Matter properties, both from
particle physics and galactic halo physics, will be demonstrated.

\end{abstract}
\maketitle

\section{Introduction}

\begin{figure}[t]
\begin{center}
\includegraphics[scale=0.15,angle=90]{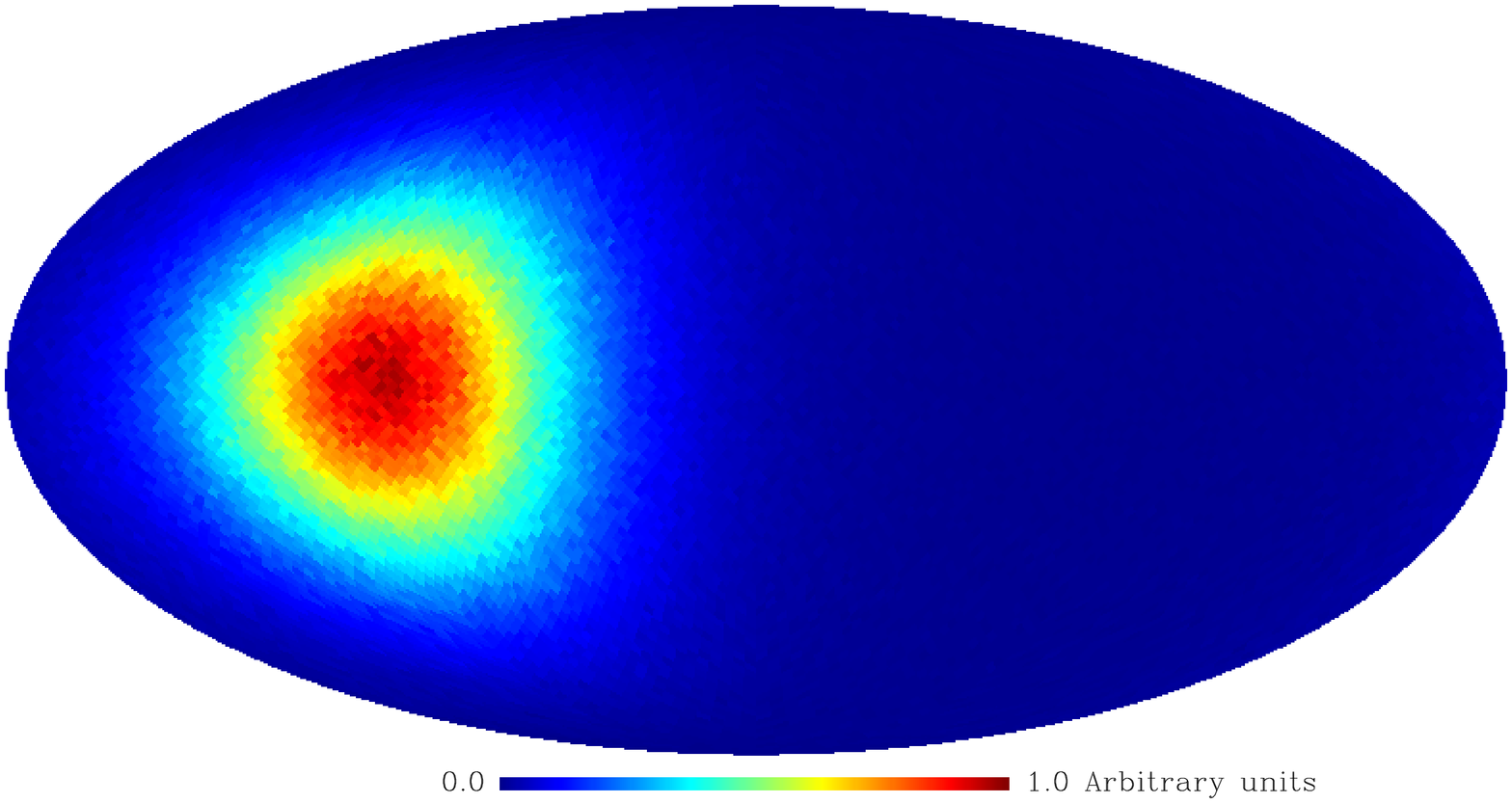}
\includegraphics[scale=0.15,angle=90]{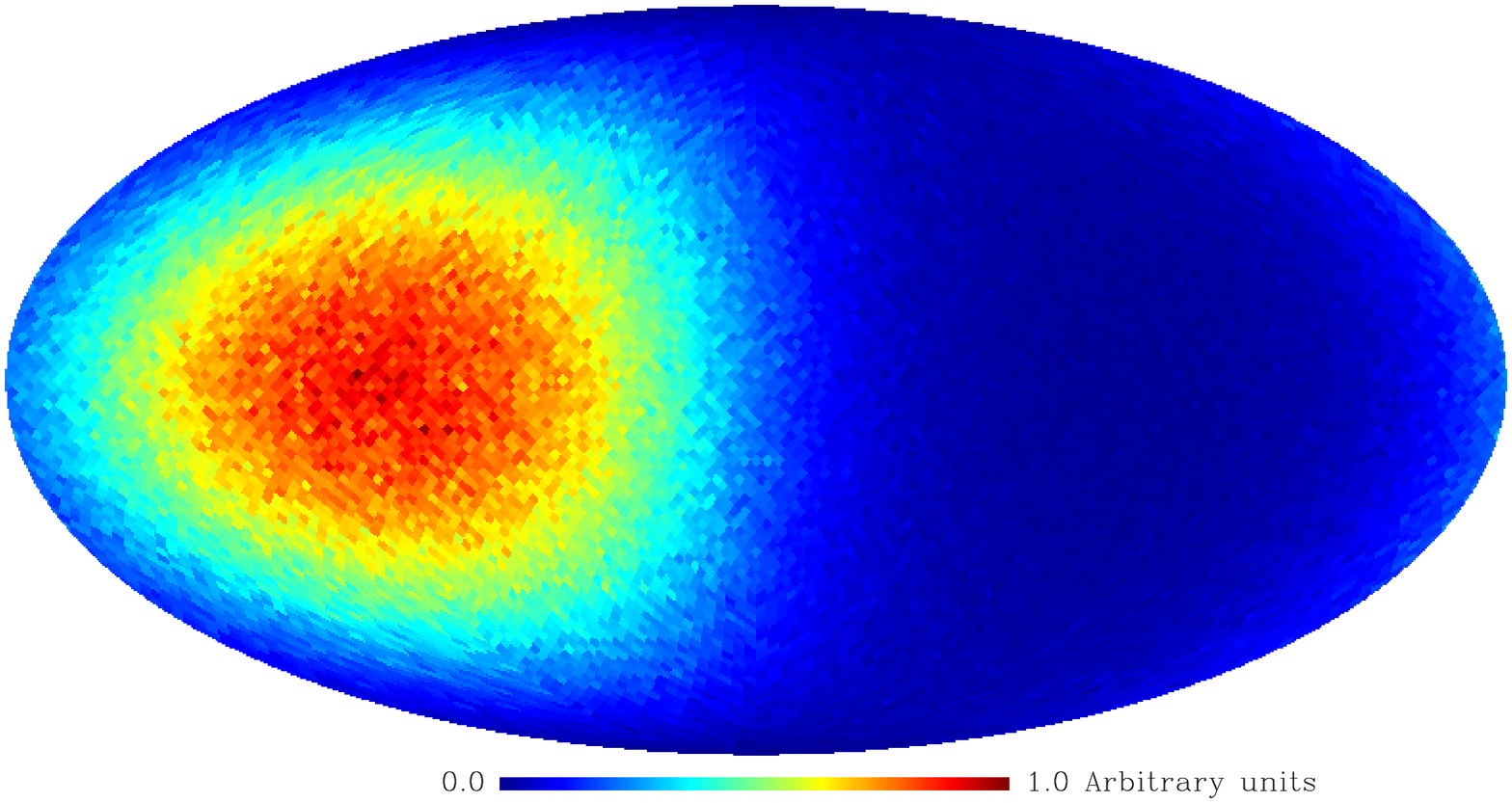}
\includegraphics[scale=0.15,angle=90]{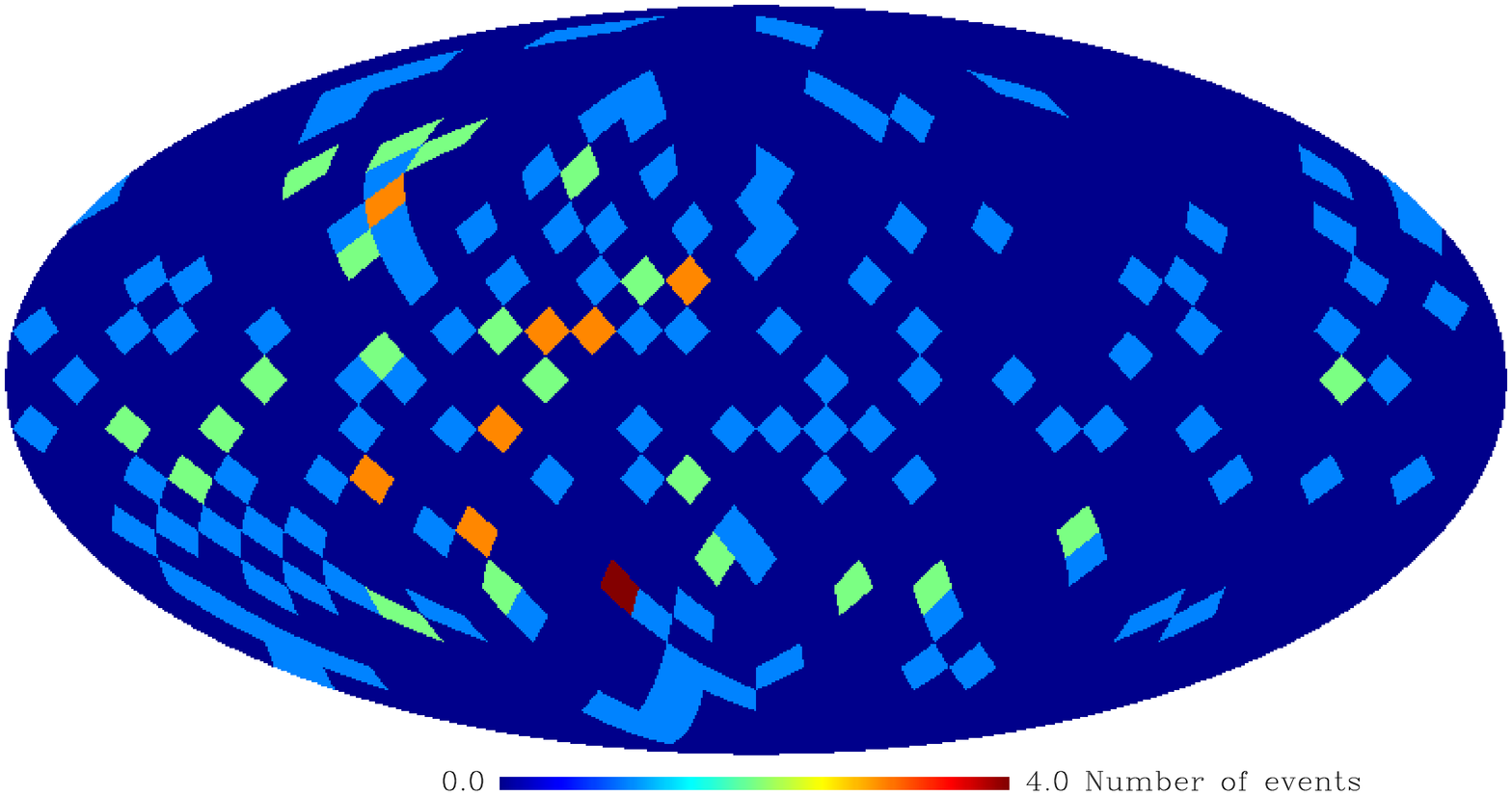}
\caption{From left to right : WIMP flux  in the case of an isothermal spherical halo,   WIMP-induced recoil distribution 
 and a typical simulated measurement :  100 WIMP-induced recoils and 100 background
events with a low angular resolution. Recoil maps are produced for a Fluorine target, a  100 GeV.c$^{-2}$ WIMP 
 and considering recoil energies in the range  5 keV $\leq E_R \leq$ 50 keV. Figures are taken from \cite{billard.disco}.}  
\label{fig:DistribRecul}
\end{center}
\end{figure}

Taking advantage on the astrophysical framework, directional detection of Dark Matter is an interesting strategy to distinguish
 WIMP events from background ones.
Indeed, like most spiral galaxies, the Milky Way is supposed to be immersed in a halo of WIMPs which outweighs the luminous component by at 
least one order of magnitude. As the Solar System rotates around the galactic center through this Dark Matter halo, WIMPs should mainly come
 from the direction to which points the
Sun velocity vector and which happens to be roughly in the direction of the Cygnus constellation ($\ell_\odot = 90^\circ,  b_\odot =  0^\circ$).
Then, a directional WIMP flux entering in any terrestrial detectors (see fig.\ref{fig:DistribRecul} left) should infer a directional
 WIMP
 induced recoil distribution pointing toward the Cygnus Constellation (see fig.\ref{fig:DistribRecul} middle).
 It corresponds to the expected WIMP signal probed by directional detectors which presents a strong anisotropy (\cite{spergel}) while the background angular distribution
  should be isotropic. Hence, we argue that a clear and unambiguous identification of a Dark matter
  detection could be done by showing the correlation of the measured signal with the direction of the solar motion.\\
 
 Several project of directional detectors are being developed (\cite{drift,newage,mimac,dmtpc,d3,emulsions}). We present a
 complete Bayesian analysis framework dedicated to directional data. The first step when analysing directional data should be to look for a signal pointing toward the
  Cygnus Constellation with a sufficiently high significance. If no evidence in favor of a Galactic origin of the signal is deduced from the previous
 analysis, then an exclusion limit should be derived. In this paper, we consider three different scenarios which are, how to set robust and competitive exclusion limits,
 how to authentify a Dark Matter detection and to estimate the significance of the latter. Eventually, it is also possible to go
 further with directional detection by identifying the properties of the WIMP particle in the case of a high significance detection. Of course, those three scenarios depend
 on the value of the unknown WIMP-nucleon cross section and on the detector sensitivity.
 In the following, unless otherwise stated, we consider the following detector characteristics: a 10 kg of CF$_4$ detector with a recoil energy range of 
 5 keV $\leq E_R \leq$ 50 keV and a data acquisition time of 3 years.

\section{Directional framework}

Directional detection depends   crucially on the WIMP velocity distribution. The isothermal sphere halo model is 
often considered but it is worth going beyond this standard paradigm in the case of model-independent analysis. 
Indeed, recent  results from 
N-body simulations are in favor of triaxial Dark Matter haloes with anisotropic velocity distributions~(\cite{nezri}). Moreover, recent observations of Sagittarius stellar
tidal stream have shown evidence for a triaxial Milky Way Dark Matter halo (\cite{Law:2009yq}).

The multivariate Gaussian WIMP velocity distribution corresponds  to the generalization of the standard isothermal sphere with a density profile 
$\rho(r)\propto 1/r^2$, leading to a smooth WIMP velocity distribution,  
a flat rotation curve and no substructure. The WIMP velocity distribution in 
the laboratory frame is   given by,
\begin{equation}
f(\vec{v}) = \frac{1}{(8\pi^3\det{\boldsymbol\sigma}^2_v)^{1/2}}\exp{\left[-\frac{1}{2}(\vec{v} - \vec{v}_{\odot})^T {\boldsymbol\sigma}^{-2}_v(\vec{v} - \vec{v}_{\odot})\right]}
\end{equation}
where ${\boldsymbol\sigma}_v = \text{diag}[\sigma_{x}, \sigma_{y}, \sigma_{z}]$ is the velocity dispersion tensor 
assumed to be diagonal in the Galactic rest frame ($\hat{x}$, $\hat{y}$, $\hat{z}$) and $\vec{v}_{\odot}$ is the Sun motion with respect to
the Galactic rest frame. When neglecting the Sun peculiar velocity and the Earth orbital 
velocity about the Sun,  $\vec{v}_{\odot}$ corresponds to the detector velocity in
the Galactic rest frame and is taken to be $v_{\odot} = 220$ km.s$^{-1}$ along the $\hat{y}$ axis pointing toward the Cygnus constellation at 
($\ell_{\odot} = 90^{\circ}$, $b_{\odot} = 0^{\circ}$). 
 The velocity anisotropy $\beta(r)$, is then defined as
  \begin{equation}
  \beta(r) = 1 - \frac{\sigma^2_{y} + \sigma^2_{z}}{2\sigma^2_x}
  \label{eq:beta}
  \end{equation}
According to N-body simulations, the $\beta$ parameter at the Solar radius   spans the range $0-0.4$, corresponding to 
radial anistropy.\\
In the following, Unless otherwise stated, we consider the standard halo model to generate simulated data, {\it i.e.} an isotropic velocity distribution ($\beta=0$)
 in which case the velocity dispersions are related to the local circular velocity $v_0 = 220$ km/s as $\sigma_{x} = \sigma_{y} = \sigma_{z} = v_0/\sqrt{2}$.\\  
  
The  directional recoil  rate  is given by  (\cite{gondolo}) :
\begin{equation}
\frac{\mathrm{d}^2R}{\mathrm{d}E_R\mathrm{d}\Omega_R} = \frac{\rho_0\sigma_0}{4\pi m_{\chi}m^2_r}F^2(E_R)\hat{f}(v_{\text{min}},\hat{q}),
\label{directionalrate}
\end{equation}
with $m_{\chi}$ the WIMP mass, $m_r$ the WIMP-nucleus reduced mass, $\rho_0=0.3 \ {\rm GeV/c^2/cm^3}$   the local Dark Matter density, $\sigma_0$   the
WIMP-nucleus elastic scattering cross section, $F(E_R)$  the form factor  (using the axial expression from (\cite{lewin})) and  
$v_{\text{min}}$ the   minimal WIMP velocity required to produce a
nuclear recoil of energy $E_R$. 
Finally, $\hat{f}(v_{\text{min}},\hat{q})$ is the three-dimensional Radon transform of the WIMP 
velocity distribution $f(\vec{v})$.  Using the Fourier slice theorem (\cite{gondolo}),    the Radon transform of the 
multivariate 
Gaussian is,
\begin{equation}
\hat{f}(v_{\text{min}},\hat{q}) = \frac{1}{(2\pi\hat{q}^T{\boldsymbol\sigma}^2_v\hat{q})^{1/2}}\exp{\left[-\frac{\left[v_{\text{min}} - \hat{q}.\vec{v}_{\odot}\right]^2}{2\hat{q}^T{\boldsymbol\sigma}^2_v\hat{q}}\right]}.
\end{equation}

 \section{Case of a null detection}

We present a Bayesian estimation of exclusion limits dedicated to directional data where only the angular part of the event distribution is
considered \cite{billard.exclusion}. The fact that both signal and background angular spectra are well known allows to derive upper limits using the Bayes' theorem. 
Considering an extended likelihood function with flat priors for both the expected number of WIMP events ($\mu_s$) and background events ($\mu_b$), and taking the evidence as  a normalization factor, it is reduced to 
  \begin{equation}
 \mathscr{L}(\vec{\theta}) = \frac{(\mu_s + \mu_b)^N}{N!}e^{-(\mu_s + \mu_b)} \ \times \ \prod_{n = 1}^{N} \left[ \frac{\mu_s }{\mu_s + \mu_b} S(\vec{R}_n)  + \frac{\mu_b }{\mu_s + \mu_b}B(\vec{R}_n)\right ]
 \end{equation}
 where $\vec{\theta} = \{ \mu_{s}, \mu_{b}\}$, $\vec{R}_n$ refers to the characteristics of the events, direction and
energy and $N$ corresponds to the total number of observed events. Hence, the probability density function of the parameter of interest $\mu_s$ can be
 derived by marginalizing $\mathscr{L}(\vec{\theta})$ over the parameter $\mu_b$. The excluded number of WIMP events $\mu_{\rm exc}$, corresponding to an excluded
 cross-section, at 90\% CL is obtained by solving:
\begin{equation}
\int_0^{\mu_{\rm exc}} \mathscr{L}(\mu_{s}) \ d\mu_s = 0.9,
\end{equation}

\begin{figure}[t]
\begin{center}
\includegraphics[scale=0.29,angle=0]{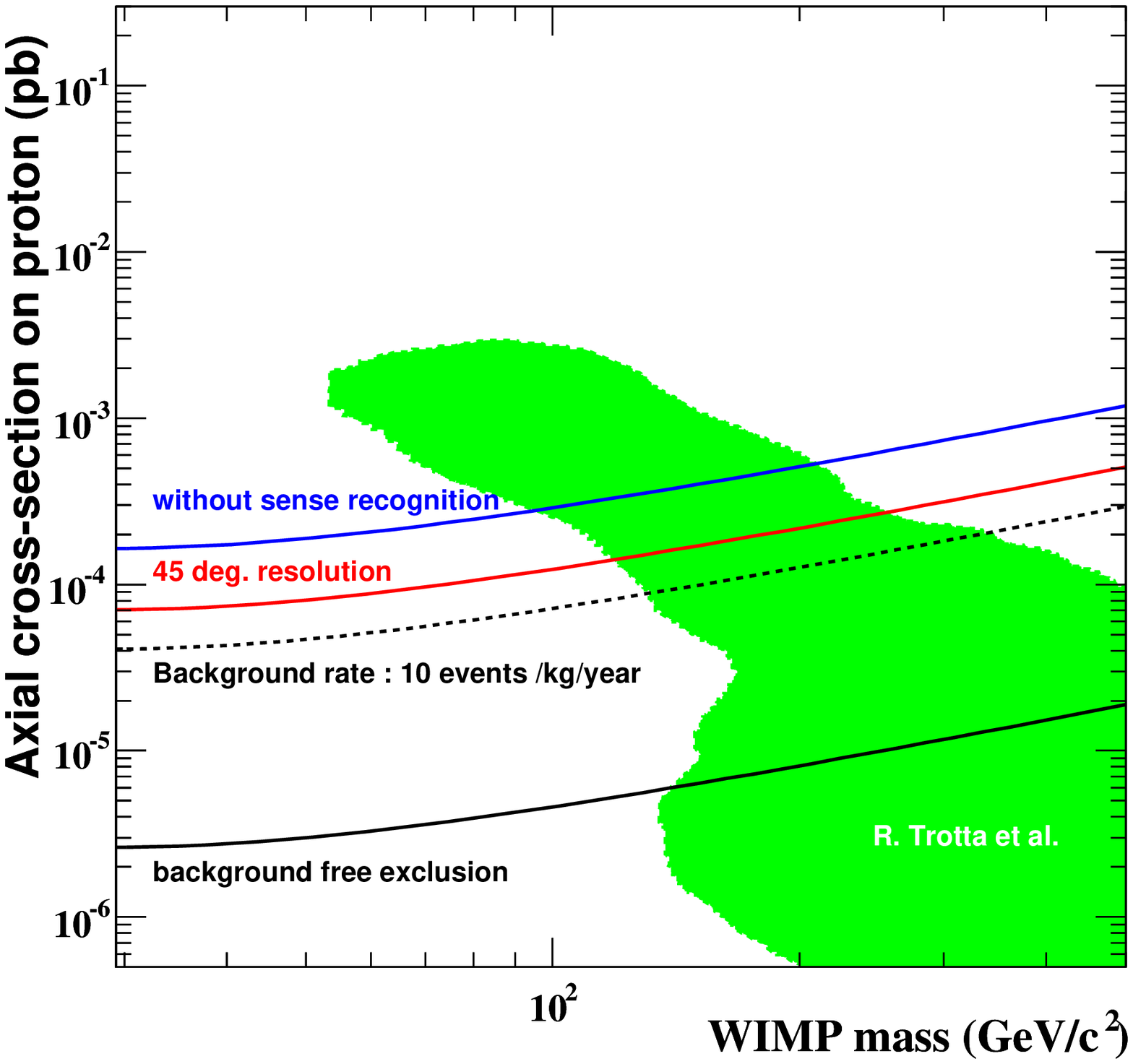}
\hspace{5mm}
\includegraphics[scale=0.29,angle=0]{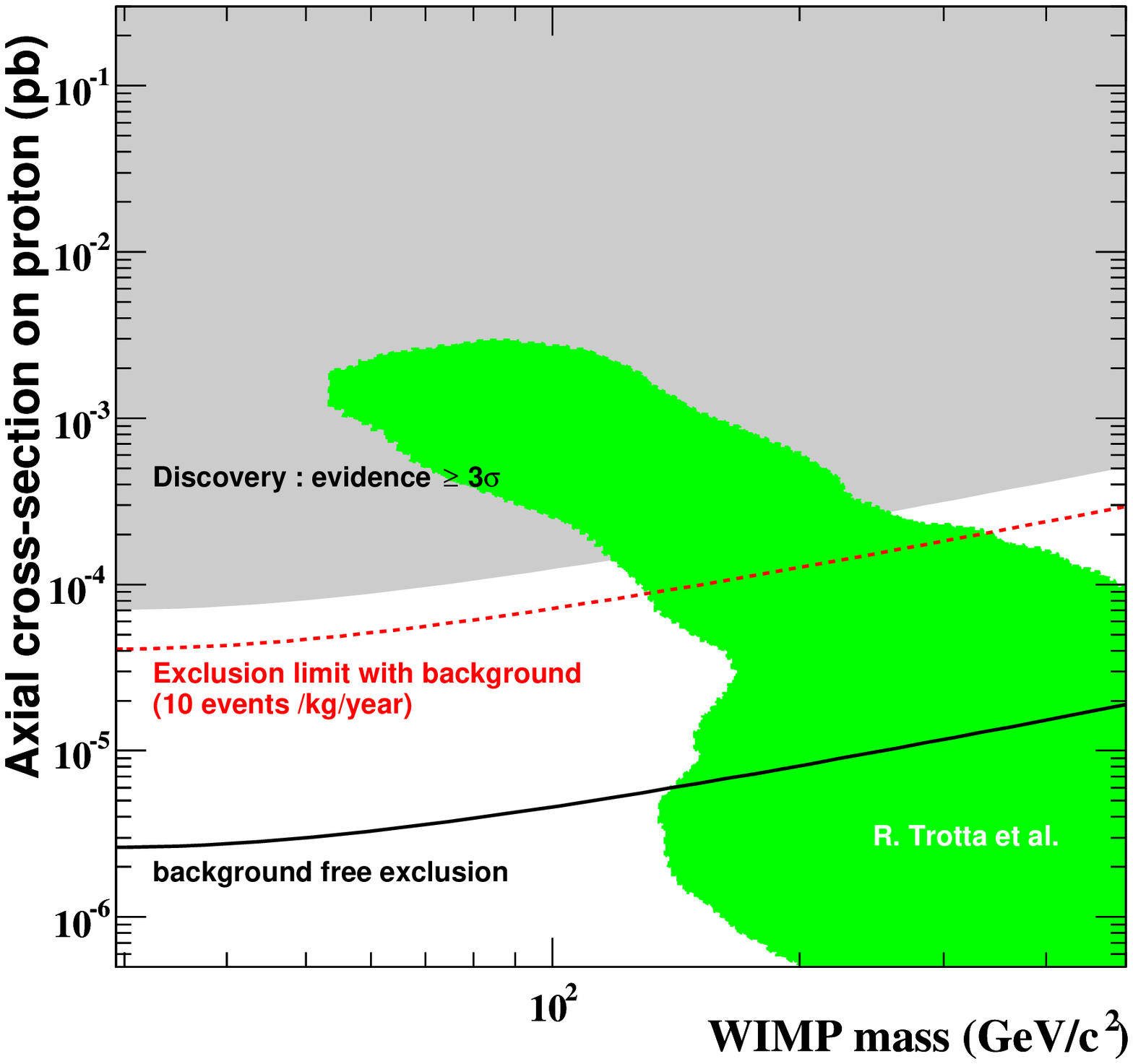}
\caption{Spin dependent cross section on proton (pb) as a function of the WIMP mass ($\rm GeV/c^2$), in the     pure-proton approximation. The green shaded area represent 
the favored region obtain in the constrained minimal supersymmetry. Left:  
projected exclusion limits of a forthcoming 
directional detector of 30 kg.year are presented in four cases : 
background-free (black solid line), with a background rate of 10 events/kg/year with sense recognition and perfect angular resolution (black dotted line), with the same
background rate considering $\sigma_{\Theta} = 45^{\circ}$ (red solid line) and considering no sense recognition (blue solid line). Right: the shaded area presents the
 $3 \sigma$  discovery region.}  
\label{fig:BackgroundPureSanSHT}
\end{center}
\end{figure}

For each detector configuration and input, we have used 10 000 toy Monte
Carlo experiments in order to evaluate the frequency distributions of the excluded cross-section. Then, from each distribution, we can derive the median value of the 
excluded cross-section $\rm \sigma_{med}$. More details may be found in (\cite{billard.exclusion}). In the following, we have considered the effect
 of three experimental issues on exclusion limits using directional detection: background contamination, angular resolution and
the sense recognition capability. Indeed, even though several progresses have been done, these experimental issues remain challenging for
directional detection of Dark Matter. On the left panel of figure \ref{fig:BackgroundPureSanSHT}, we present the exclusion limits in the $(m_{\chi},\sigma_{n})$ plane 
for four different cases: background free (black solid line), ideal detector with a background contamination of 10 events/kg/year (black dotted line), the effect of a
$45^{\circ}$ degree angular resolution (red solid line) and the effect of no sense recognition (blue solid line). It should be noticed that in the case of angular resolution
and sense recognition, a background rate of 10 events/kg/year is also considered.\\
From the left panel of figure \ref{fig:BackgroundPureSanSHT}, it can be seen that the main experimental issues when setting exclusion limits with directional detection is obviously 
the background contamination. However, with a background contamination of 300 expected events, if one do not use directional information (Poisson limit),
 the exclusion limit should be two
order of magnitude above the one corresponding to the background free configuration. Then, we can deduce from the left panel of figure \ref{fig:BackgroundPureSanSHT}
 that directional
information allows us to improve exclusion limits by about one order of magnitude, highlighting the interest of this kind of direct detection.\\
Without sense recognition, a recoil coming from ($\cos\gamma$,$\phi$) cannot be distinguished from a recoil coming
from ($-\cos\gamma$,$\phi + \pi$) leading to an expected angular distribution, from WIMP events, less anisotropic and then closer to the one from background events.
One can see from the left panel of figure \ref{fig:BackgroundPureSanSHT} that the effect of having or not the sense recognition capability in the case of high background 
contamination will only affect the exclusion limit by a factor of $\sim$ 4.
 Taken at face value, this result suggests that sense recognition may not be so important for directional detection when setting exclusion limits.\\
Having a finite angular resolution means that a recoil initially coming from the direction  $\hat{r}(\theta, \phi)$ is 
reconstructed as a recoil ${\hat{r}}^{\, \prime}(\theta^\prime, \phi^\prime)$ with a gaussian dispersion of 
 width $\sigma_\Theta$. Then, the effect of angular resolution is that the angular distribution is smoother and hence degrades the discrimination between the expected WIMP
 and background events. The effect of an angular resolution of $\sigma_\Theta = 45^{\circ}$ at high background contamination can be seen by comparing 
 the red solid line and the black
dotted one from the left panel of figure \ref{fig:BackgroundPureSanSHT}. One can see that an angular resolution of $45^{\circ}$ only degrade the exclusion limit of a factor
$\sim$ 2 for a
background contamination of 10 events/kg/year.
Hence, as far as exclusion limits 
 are concerned, the effect of angular resolution is relatively small.\\ 
 
 To conclude this study, one can see from the left panel of figure \ref{fig:BackgroundPureSanSHT} that directional detection should be able to reach a large fraction of
 some supersymmetric model, highlighting the need for this kind of direct detection of Dark Matter.

\section{Case of positive detection}

An observed recoil map such as the one on the right of figure \ref{fig:DistribRecul} could be obtained with a 10 kg CF$_4$ detector with
a WIMP-nucleon cross-section of $\sigma_n = 1.5\times10^{-3}$ pb and with a background rate of $\sim 0.07$ kg$^{-1}$.day$^{-1}$ in $\sim 5$ months exposition time.
 At first
sight, it seems difficult to conclude from this simulated recoil map that it does contain 
a fraction of WIMP events pointing toward the direction of the Solar motion. 
A likelihood analysis is developed (\cite{billard.disco}) in order to retrieve from a recoil map : the main direction of the incoming events in 
Galactic coordinates ($\ell, b$) and the number of WIMP events contained in the map. The likelihood value is estimated using a binned map 
of the overall sky with  Poissonian statistics,  as follows :
 \begin{equation}
 \mathscr{L}(m_\chi,\lambda, \ell,b) = \prod_{i=1}^{N_{\rm pixels}} P( [(1-\lambda) B_i + \lambda S_i(m_\chi ;\ell,b) ]|M_i)
 \end{equation}
where $B$ is the  background spatial distribution 
taken as isotropic, $S$ is the WIMP-induced recoil distribution and $M$ is the measurement. 
This is a four parameter likelihood analysis with $m_\chi$, 
 $\lambda = \mu_s/(\mu_s+\mu_b)$ the  WIMP fraction (related to the  background 
rejection power of the detector) and the coordinates ($\ell$, $b$) referring to the maximum of the 
WIMP event angular distribution.
Hence, $S(m_\chi;\ell,b)$ corresponds to a rotation of the $S(m_\chi)$ distribution 
by the angles ($\ell' = \ell - \ell_\odot$, $b' = b - b_\odot$). A scan of the four parameters with flat priors, allows to evaluate the likelihood between the measurement 
(fig.~\ref{fig:DistribRecul} right) and the theoretical distribution made of a superposition of 
an isotropic background and a pure WIMP signal (fig. \ref{fig:DistribRecul} middle). By scanning on $\ell$ and $b$ values, we ensure 
that there is no prior on the direction of the center of the WIMP-induced recoil distribution. As the observed map  is considered as a superposition of  
the background and the WIMP signal distributions, no assumption on the origin of each event is needed. Moreover, the likelihood method allows to recover $\lambda$,
 the WIMP fraction contained in the data.\\

\begin{figure}[t]
\begin{center}
\includegraphics[scale=0.25,angle=270]{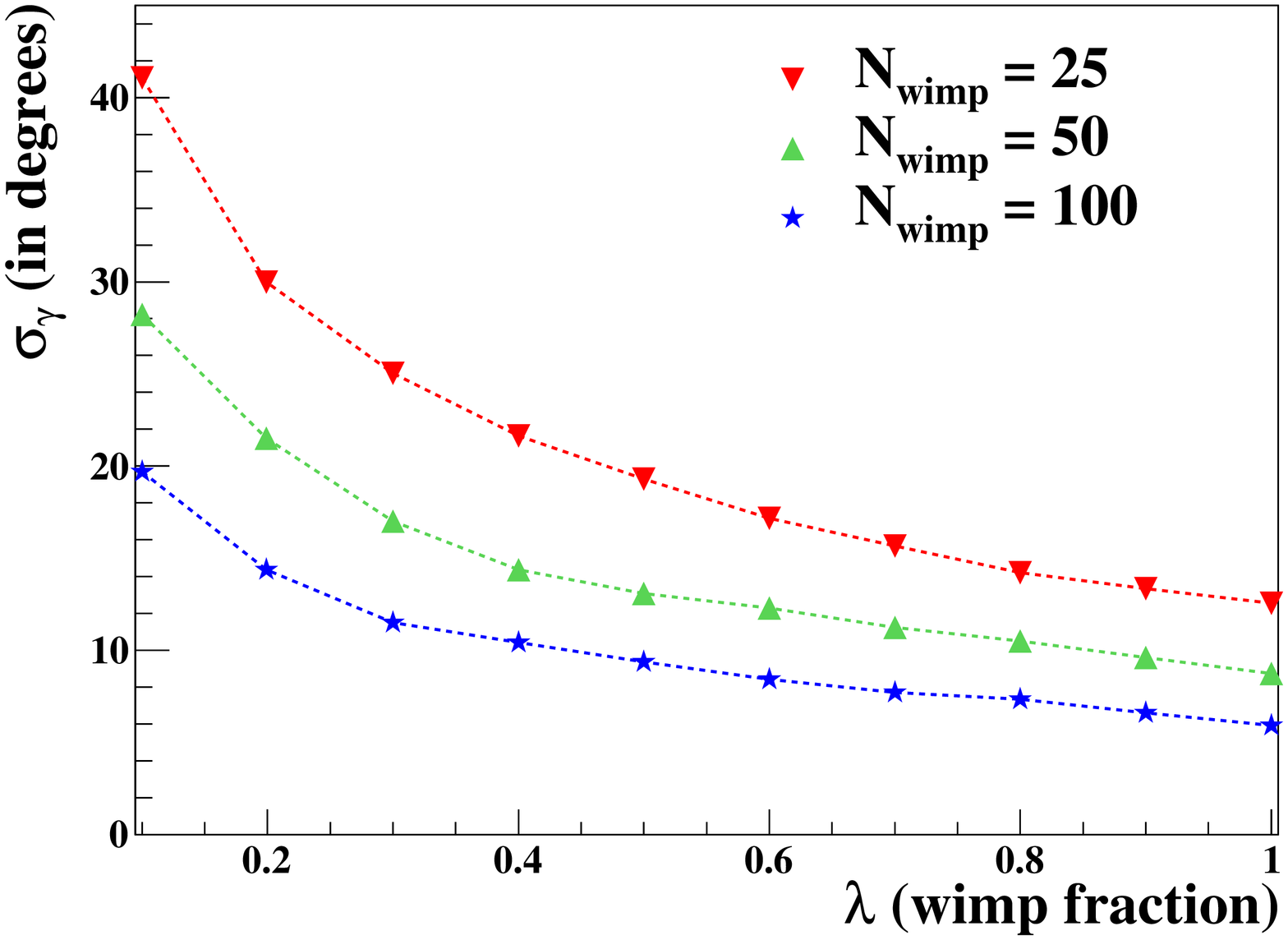}
\hspace{5mm}
 \includegraphics[scale=0.25,angle=270]{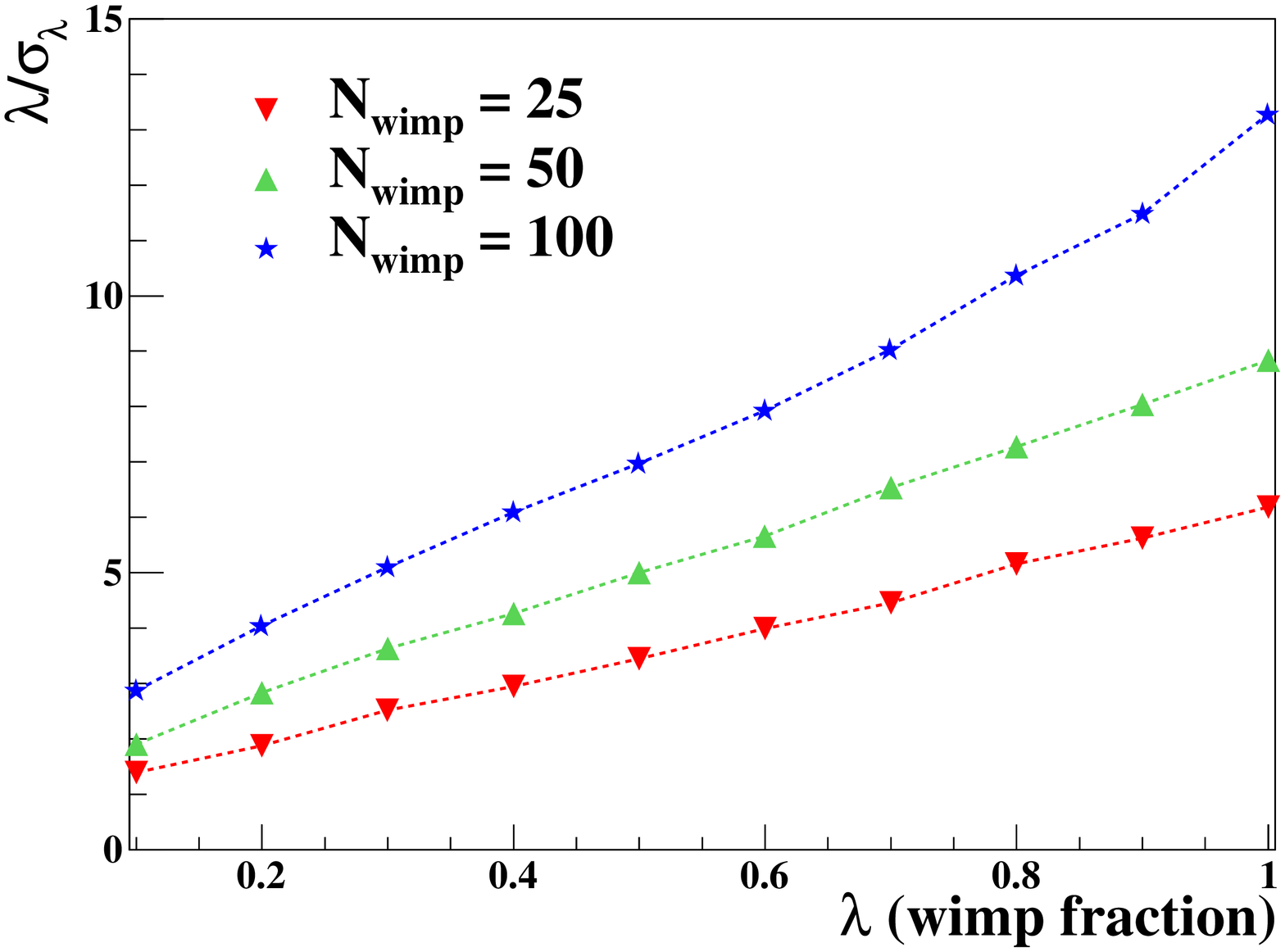}
\caption{Left panel presents the directional signature $\sigma_{\gamma}$ (in degrees) as a function of the WIMP fraction. 
Right panel presents the significance ($\lambda/\sigma_\lambda$) as a function of the WIMP fraction $\lambda= \mu_{s}/(\mu_s+\mu_b)$. 
Results are produced for a $^{19}$F target, a  100 GeV.c$^{-2}$ WIMP 
 and considering recoil energies in the range  5 keV $\leq E_R \leq$ 50 keV. Figures are taken from \cite{billard.disco}.}  
\label{fig:exposition}
\end{center}
 \end{figure}

In order to explore the interest of directional detection combined with such likelihood analysis, we have done some systematical studies 
using $10^4$ experiments for various number of WIMP events ($N_{\rm wimp}$) 
and several values of WIMP fraction in the observed map ($\rm \lambda$), ranging from 0.1 to 1. 
For a given cross-section, these two parameters are 
related respectively with the exposure and the rejection power of the offline analysis 
preceding  the likelihood method.\\
Figure~\ref{fig:exposition} presents on the left panel the directional signature, taken as the value of $\sigma_{\gamma} = \sqrt{  \sigma_\ell
\sigma_b }$,  the radius of the  $68 \%$ CL contour of the marginalised
$\mathscr{L}(\ell,b)$ distribution, as a function of $\lambda$. It is related to the ability to 
recover the main signal direction and to sign its Galactic origin. It can be  noticed that the directional  
signature is of the order of 10$^\circ$ to 20$^\circ$ on a wide range of WIMP fractions.
Even for low number of WIMPs and for a low WIMP fraction (meaning a poor rejection power), 
the directional signature remains clear. From this, we conclude
that a directional evidence in favor of Galactic Dark Matter may be obtained with 
upcoming experiments even at low exposure and with a non-negligible background 
contamination.\\
However, a convincing proof of the detection of WIMPs would require  a directional  
signature with  sufficient significance. We defined the significance of this identification strategy as $\lambda/\sigma_\lambda$, presented
on figure~\ref{fig:exposition} (right panel) as a function of $\lambda$. As expected, the significance is increasing both with the 
number of WIMP events and with the WIMP fraction, but we can notice that an evidence ($\rm 3 \sigma$) or a discovery ($\rm 5
\sigma$) of a Dark Matter signal would require either a larger number of WIMPs or a lower background contamination.\\
To conclude, on the right panel of figure \ref{fig:BackgroundPureSanSHT} we have presented the area in the $(m_{\chi},\sigma_{n})$ plane for wich a directional detector, with
30 kg.year exposure, could reach a $3\sigma$ significance detection in average. Then, a directional detector as the one considered here, should be able to reach a 
$3\sigma$ significance detection of Dark Matter down to a WIMP-nucleon cross section of about $10^{-4}$ pb.

 \section{Identification of Dark Matter}

As we have seen previously, directional detection should provide powerful arguments in order to authentify a Dark Matter detection, or to set
robust and competitive exclusion limits in the case of a null detection. However, it is possible to go further by exploiting all the information
 from a directional detector,
{\it i.e.} the energy and the direction of each event. We will then consider the most optimistic scenario where the WIMP nucleon cross-section is sufficiently large to get a
high significance Dark Matter detection. Then, we show for the first time the possibility to 
constrain the  WIMP properties, 
both from particle physics ($m_\chi, \sigma_n$) and galactic Dark Matter halo physics (velocity dispersions) (\cite{billard.ident}). 
This leads to an identification of non-baryonic Dark Matter, which could be reached within few years by upcoming directional detectors (\cite{white}).
The model is characterized by 8 free parameters which are 
$\{m_{\chi}, \log_{10}(\sigma_n), l_{\odot}, b_{\odot},\sigma_{x}, \sigma_{y}, \sigma_{z}, R_b\}$, 
where the direction $(l_{\odot}, b_{\odot})$ refers to 
the main direction of the recorded events (see sec.4), $\sigma_n$ is the WIMP-nucleon cross section directly related to $\sigma_0$ in the 
pure proton approximation and $R_b$ is the background rate.
We have considered flat prior for each parameter. In such case, the Bayes'
 theorem is simplified and the target distribution   reduces to a   8 dimensional 
 likelihood function $\mathscr{L}(\vec{\theta})$ dedicated to unbinned data as given by equation (3.1). The latter is sampled by using a MCMC analysis 
 based on the Metropolis-Hastings algorithm, using chain 
 subsampling   according to  the burn-in and  correlation lengths to deal only with 
 independent samples. More details may be found in (\cite{billard.ident}).

\begin{figure*}[t]
\begin{center}

\includegraphics[scale=0.30,angle=0]{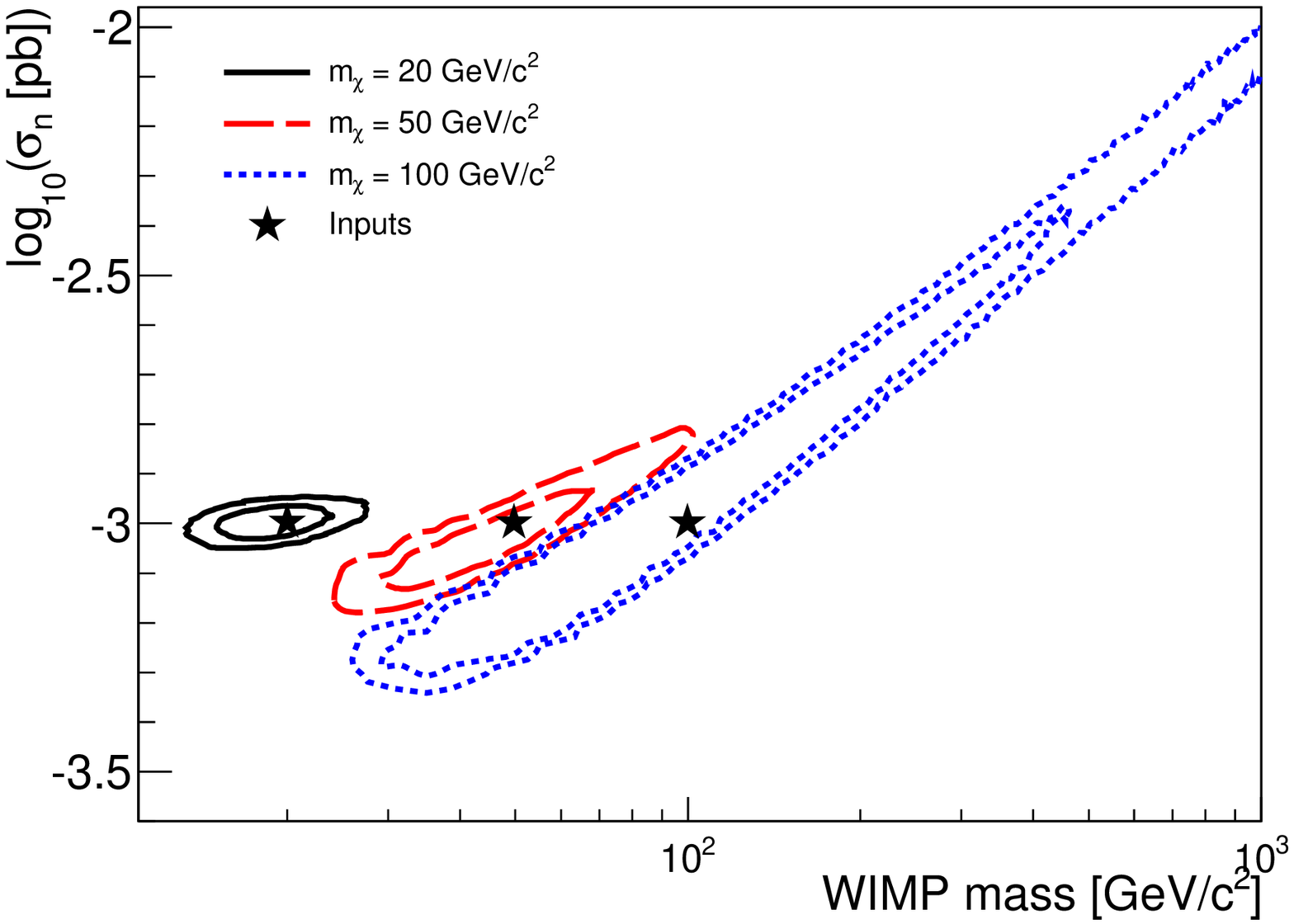}
\includegraphics[scale=0.30,angle=0]{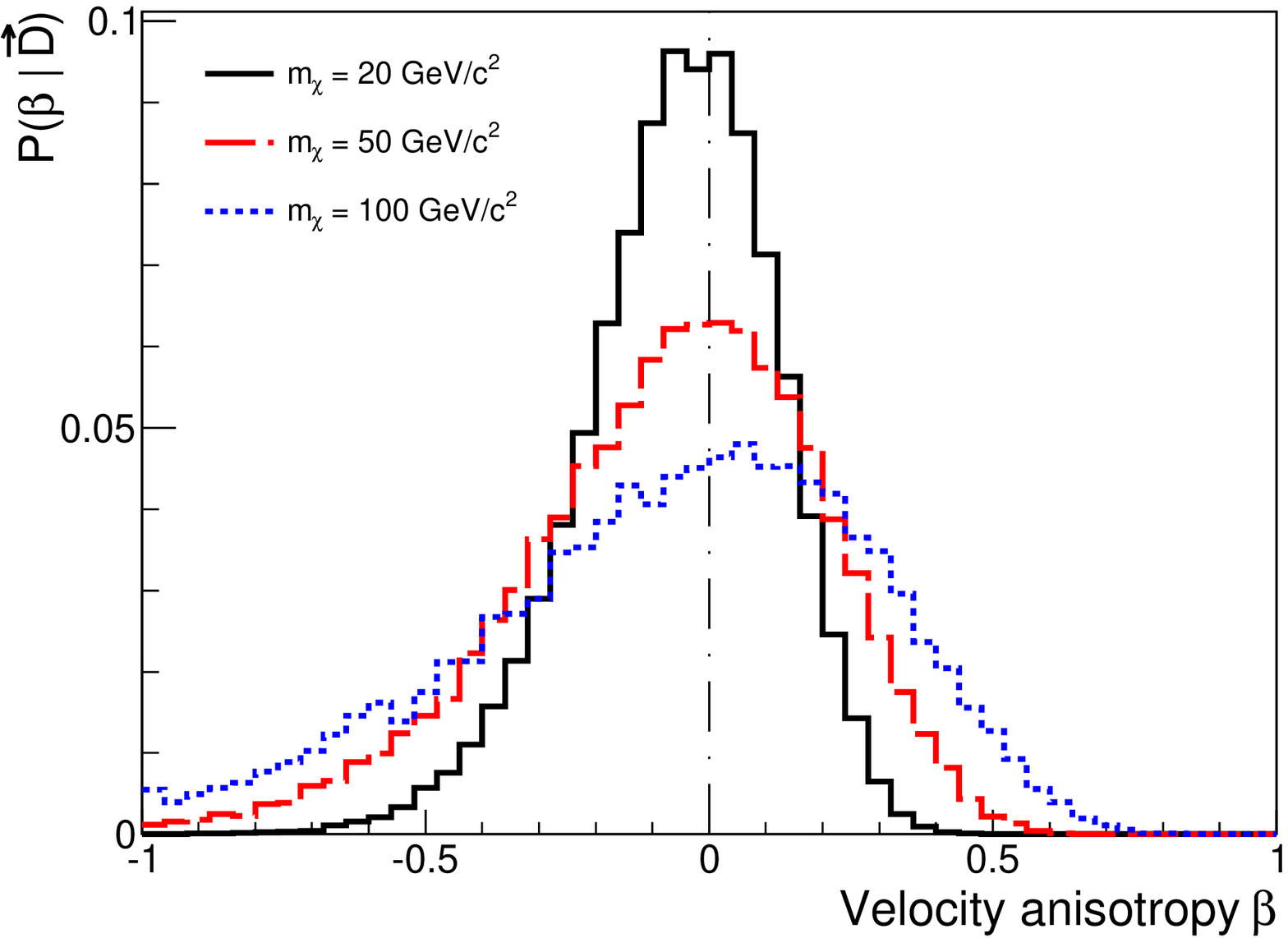}
\caption{Left panel : 68\% and 95\% contour level in the ($m_{\chi},\sigma_n$) plan, for the isotropic input model and for a WIMP mass 
equal to 20, 50 and 100 $\rm GeV/c^2$. 
Right panel : posterior PDF distribution of the $\beta$ parameter for the same models. Figures are taken from \cite{billard.ident}.} 
\label{fig:WIMPMass}

\includegraphics[scale=0.30,angle=0]{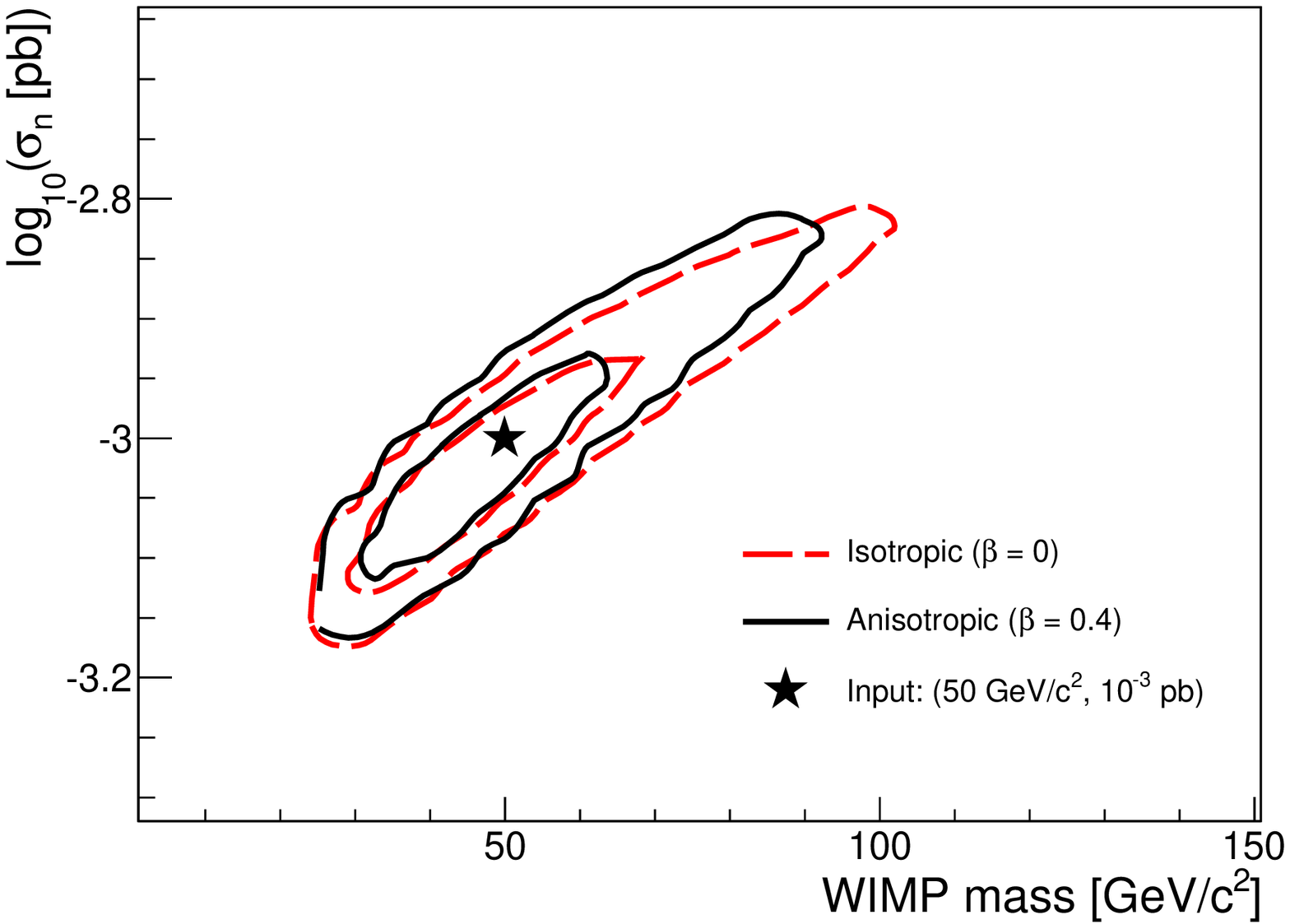}
\includegraphics[scale=0.30,angle=0]{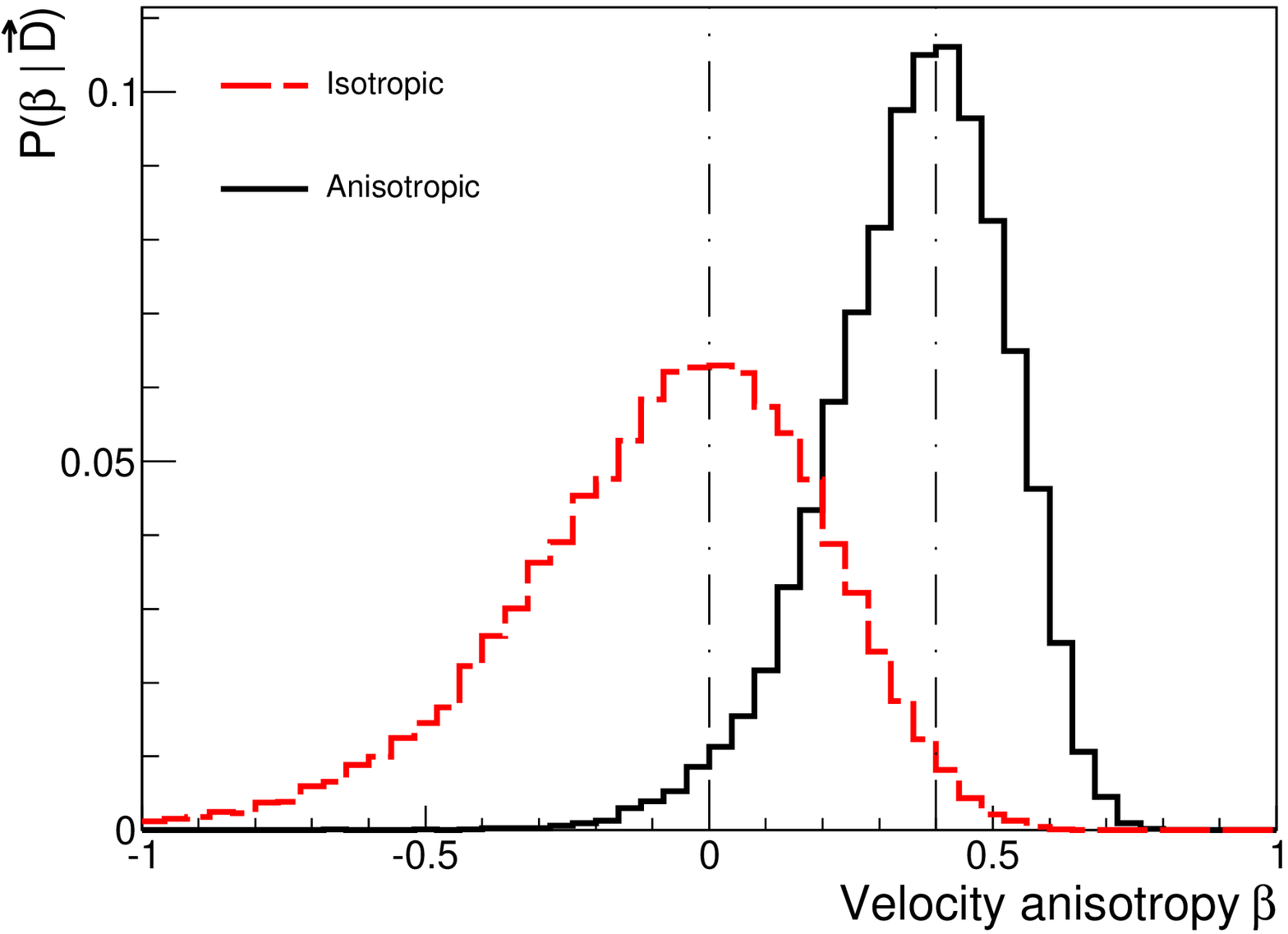}
\caption{Left panel : 68\% and 95\% contour level in the ($m_{\chi},\sigma_n$) plan, for a 50 $\rm GeV/c^2$ WIMP and for two input models : isotropic ($\beta=0$) and 
triaxial ($\beta=0.4$). 
Right panel : posterior PDF distribution of the $\beta$ parameter for the same models. Figures are taken from \cite{billard.ident}.}  
\label{fig:HaloT4}
\end{center}
\end{figure*}

In the following, we briefly
discuss the effect of some of the input parameters on the different constraints in order to estimate the performance of this analysis tool. 
We first focus on the impact of the input WIMP mass. To do so, we have simulated
three different sets of directional data corresponding to an input WIMP mass of 
$m_{\chi} = 20, 50, 100 \ {\rm GeV/c^2}$ with a constant WIMP-nucleon cross-section  $\sigma_n = 10^{-3} \ {\rm pb}$ and the standard isotropic halo model. 
The results from the three MCMC runs are illustrated on figure \ref{fig:WIMPMass}. 
We present for the three WIMP masses, on the left panel,  the 68\% and 95\% CL contours in the 
($m_{\chi},\log_{10}(\sigma_n)$) plan and on the right panel, the posterior PDF $P(\beta|\vec{D})$ of 
the anisotropy velocity parameter $\beta$.\\
It can be deduced from figure~\ref{fig:WIMPMass} that the constraints strongly depend on the input WIMP mass, but in each case,
 the constraints are consistent with the input values. Then, this analysis has been shown to be working for any input WIMP mass although the constraints are stronger for
 light WIMPs. This is due to the fact that the signal characteristics, {\it i.e} the slope of  the 
energy distribution and the width of the angular distribution, evolve slowly with the WIMP mass once $m_{\chi} \geq 100$ GeV/c$^2$, as shown in (\cite{billard.disco}).\\
In the following, we investigate the effect of an extremely triaxial halo model with $\beta=0.4$ ({\it i.e.} $\sigma_x = 200$ km/s; $\sigma_z = 169$ km/s; $\sigma_y = 140$ km/s)
on the estimation   of the Dark Matter parameters ($m_{\chi},\sigma_n,\beta$). The results from the MCMC run on a simulated 
dataset corresponding to a WIMP mass of 50 GeV/c$^2$ with the anisotropic halo model are presented on
figure \ref{fig:HaloT4}. For convenience and comparison, 
the results from a benchmark input model (isothermal sphere with a 50 GeV/c$^2$ WIMP) are recalled.\\
From the left panel of figure \ref{fig:HaloT4}, we can conclude that the two halo models give similar constraints which are both consistent with the input values.
 In fact, and as foreseen, the fact that the velocity
dispersions are set as free parameters in the MCMC analysis allows to avoid induced bias due to wrong model assumption. 
From the right panel of figure \ref{fig:HaloT4} we can deduce that the $\beta$ parameter is well constrained: 
$\beta = 0.38^{+0.2}_{-0.1}$ and strongly in favor of an anisotropic Dark Matter halo.\\
As one can see, directional detection should provide strong constraints on Dark Matter both from particle physics (mass and cross section) and also from astrophysics. Indeed,
the fact that it is possible to constrain the local WIMP velocity distribution gives the opportunity to estimate the Dark Matter halo profile and start a Dark Matter
astronomy.

\section{Conclusion}
 
As a conclusion, it can be highlighted that Directional detection should provide unambiguous arguments in favor of a positive or a null Dark Matter detection.
Indeed, the angular distribution of the expected WIMP events is very unlickely to be mimicked by known backgrounds. That way, we have shown that Bayesian analysis of
directional data are suitable to either set exclusion limits in the case of a null detection or to clearly authentify a positive detection. Moreover, we have seen that in the
case of a high significance detection of Dark Matter, directional detection should be able to probe the structure of the local WIMP velocity distribution and to constrain the WIMP
mass and cross section at the same time.


\end{document}